\documentclass[letterpaper, 10 pt, conference]{ieeeconf}

\IEEEoverridecommandlockouts
\usepackage{cite}
\usepackage{amsmath,amssymb,amsfonts}

\usepackage{amsthm}
\usepackage{graphicx}
\usepackage{textcomp}
\usepackage{xcolor}
\usepackage{booktabs}
\usepackage{multirow}
\usepackage{array}
\usepackage{url}
\usepackage{microtype}
\let\labelindent\relax
\usepackage{enumitem}
\usepackage{algorithm}
\usepackage{algpseudocode}
\usepackage{caption}
\usepackage{subcaption}
\usepackage[hidelinks]{hyperref}

\newenvironment{IEEEkeywords}{\par\vspace{0.5ex}\noindent\textit{Index Terms---\ignorespaces}}{\par}

\def\BibTeX{{\rm B\kern-.05em{\sc i\kern-.025em b}\kern-.08em
    T\kern-.1667em\lower.7ex\hbox{E}\kern-.125emX}}

\newcommand{\R}{\mathbb{R}}
\newcommand{\E}{\mathbb{E}}
\newcommand{\N}{\mathcal{N}}

\newcommand{\at}{\sqrt{\bar{\alpha}_t}}
\newcommand{\st}{\sqrt{1-\bar{\alpha}_t}}

\theoremstyle{plain}
\newtheorem{proposition}{Proposition}

\begin{document}

\title{CFGComposer: Compositional Audio-Text-to-Motion\\
       Generation via Copula Score Correction}

\newcommand{\AuthorOne}{Pengyang Shi}
\newcommand{\AuthorTwo}{Jinlu Zhang}
\newcommand{\AuthorThree}{Chunze Lin}
\newcommand{\AuthorFour}{Yizhou Wang}
\newcommand{\AuthorFive}{Li Yi}

\newcommand{\AffiliationOne}{Tsinghua University}
\newcommand{\AffiliationTwo}{Galbot}
\newcommand{\AffiliationThree}{CFCS, School of Computer Science, Peking University}
\newcommand{\AffiliationFour}{Shanghai Qi Zhi Institute}
\newcommand{\AffiliationFive}{State Key Laboratory of General Artificial Intelligence}

\author{%
  \authorblockN{%
    \AuthorOne\textsuperscript{1,2},
    \AuthorTwo\textsuperscript{2,3},
    \AuthorThree\textsuperscript{2},
    \AuthorFour\textsuperscript{3,5},
    \AuthorFive\textsuperscript{1,2,4}%
  }
  \authorblockA{%
    \textsuperscript{1}\AffiliationOne \qquad
    \textsuperscript{2}\AffiliationTwo\\
    \textsuperscript{3}\AffiliationThree\\
    \textsuperscript{4}\AffiliationFour \qquad
    \textsuperscript{5}\AffiliationFive%
  }
}

\hypersetup{
  pdftitle={CFGComposer: Compositional Audio-Text-to-Motion Generation via Copula Score Correction},
  pdfauthor={\AuthorOne; \AuthorTwo; \AuthorThree; \AuthorFour; \AuthorFive}
}

\makeatletter
\let\@oldmaketitle\@maketitle
\renewcommand{\@maketitle}{\@oldmaketitle
  \centering
  \includegraphics[width=0.96\textwidth]{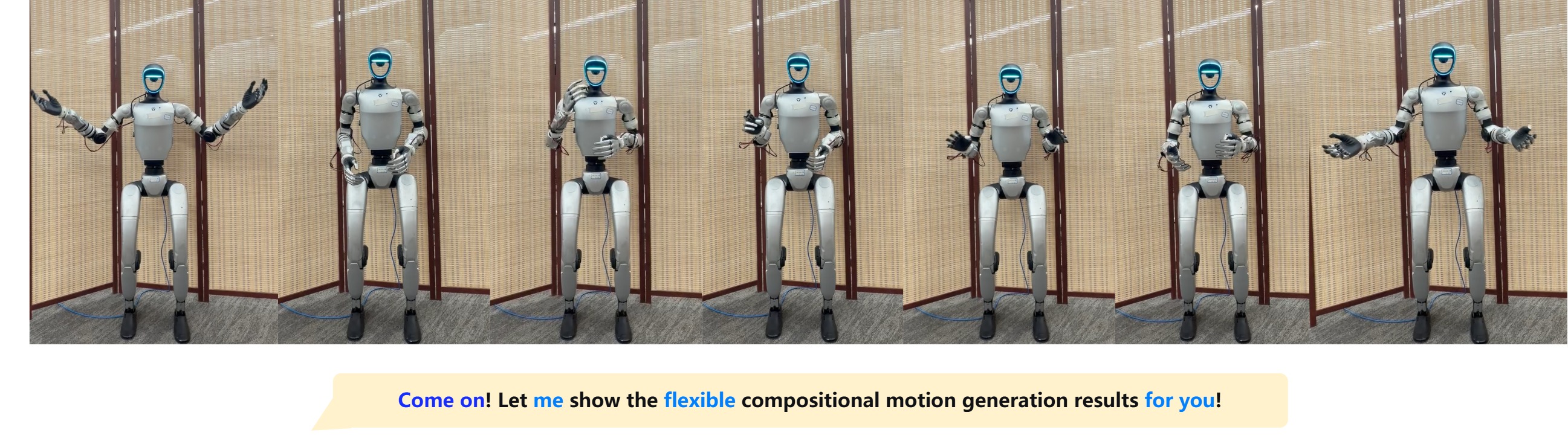}
  \captionof{figure}{\textbf{Compositional AT2M via CFG-Guided Interaction Correction.}
    Given speech audio $a$, text caption $c$ and frame window $m$, the goal is to generate
    whole-body motion that is rhythmically locked to the audio
    and semantically specific to the text in the gesture window.
   }
  \label{fig:teaser}
  \vspace{-0.4cm}
}
\makeatother

\maketitle
\thispagestyle{empty}
\pagestyle{empty}

\setcounter{figure}{1}

\begin{abstract}
Generating whole-body gestures that are simultaneously rhythmically
synchronized with speech audio and semantically aligned with text descriptions
is a fundamental challenge in embodied AI and social robotics.
We tackle this \emph{Audio-Text-to-Motion (AT2M)} problem by composing two
independently-trained diffusion models at inference time and refine bias in joint inference by modeling the copula score $s_I$.
A rigorous theoretical analysis reveals that naive epsilon blending corresponds
to a Product-of-Experts formulation that omits the \emph{copula score} $s_I$:
the cross-modal interaction between audio and text that neither marginal model
can capture.
We propose \emph{CFGComposer}:
a jointly-trained AT2M diffusion model that estimates $s_I$ via a CFG analogy
and injects it as an $\epsilon$-space correction.
Experiments on BEAT demonstrate best-in-class semantic fidelity
(SRGR $0.9916$, FID $27.18$) with GT-level beat alignment
(BA $0.1826$ vs.\ GT $0.1743$), and $76$--$83\%$ win rate in blind
pairwise human evaluation against all prior methods.
Physical deployment on a Unitree~G1 humanoid confirms direct sim-to-real
transfer.
\end{abstract}


\section{Introduction}
\label{sec:intro}

When a speaker says ``the crowd \emph{rose as one}''
and sweeps both arms upward on the stressed syllable,
they are executing a gesture that is simultaneously
(a)~rhythmically locked to the prosodic peak and
(b)~semantically keyed to the meaning of rising.
Crucially, the upward sweep is a transient event on a seconds timescale:
it must begin and peak precisely as the word ``rose'' is spoken,
not half a second earlier or later.
Getting the shape right without timing is as wrong as getting
the wrong shape---the two constraints are inseparable
and must be satisfied at frame-level precision simultaneously.
For embodied AI and social robotics, replicating this dual property is fundamental:
a humanoid whose gestures are rhythmic but semantically vacant fails the basic social test of human-likeness~\cite{mcneill1992hand}.

Many recent works address this audio-text-to-motion (AT2M) setting and pursue multimodal fusion through a variety of architectural choices:
decomposed rhythm and semantic modules~\cite{zhang2024semtalk},
speech-queried semantic masking~\cite{zhang2025echomask},
retrieval augmentation from motion libraries~\cite{zhang2024sg},
and anti-inertia masking to overcome modality eclipse~\cite{robogesture2026}.
Despite their diversity, 
all these approaches share the same fundamental data bottleneck: \textbf{a contradiction between the size of existing datasets and the amount of data required by training models}.

On the one hand, existing tri-modal datasets are not large enough. Training an AT2M model requires tri-modal information -- audio, text caption with frame windows, and reference motion -- simultaneously, but in existing datasets, such as BEAT~\cite{liu2022beat}, only 5$\%$–10$\%$ of frames contain specific semantic annotations. Annotation of tri-modal data basically relies on manual work, and is inefficient and costly. Although some works use synthetic data~\cite{robogesture2026} to try to expand the scale of dataset, the artificially created data itself is not of sufficiently high quality, limiting the model’s generative capability.

On the other hand, training a tri-modal model requires more training data than training a bimodal model. This is because a tri-modal AT2M model needs to achieve alignment among text, speech, and motion, a task that is more complex than the bimodal A2M (audio-to-motion) or T2M (text-to-motion) tasks and requires more training samples to reach alignment between modalities. Previous work~\cite{zhang2025echomask,robogesture2026} observed that when data is insufficient, the model  degenerates by relying excessively on context motions and acoustic features in audio, effectively ignoring the hard-to-capture semantic guidance.

\begin{figure*}[t]
  \centering

\includegraphics[width=0.96\textwidth]{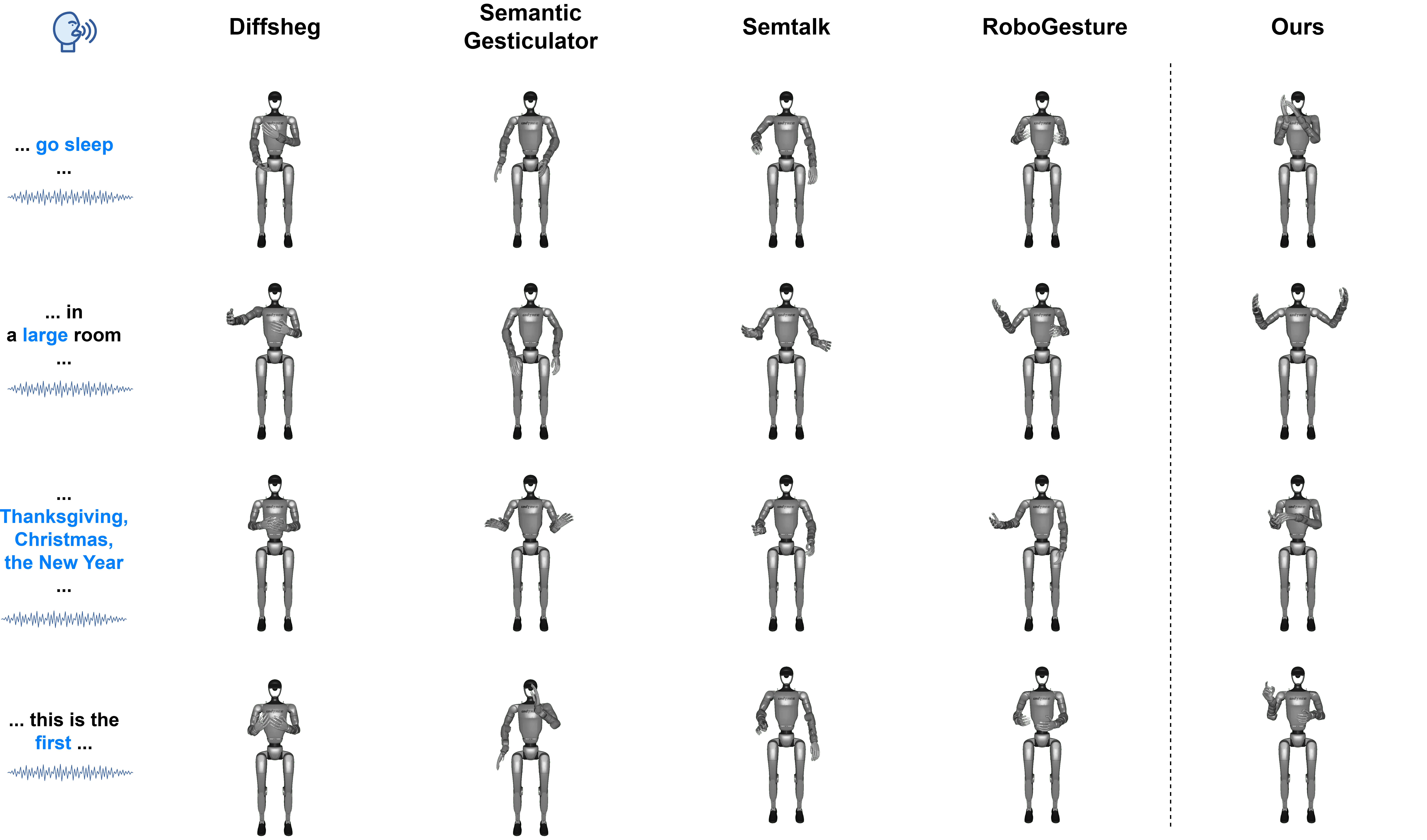}

  \caption{\textbf{Qualitative comparison on BEAT.}
    Each row shows a selected utterance with a semantic gesture annotation.
    Columns correspond to Diffsheg, SemTalk, Semantic Gesticulator,
    RoboGesture, and CFGComposer (ours).
    CFGComposer reproduces the iconic gesture shape while maintaining
    smooth human-like movement. It should be noted that in the third row, the audio input is listing festivals while the text caption is ``gesture for giving examples", CFGComposer is the only method to generate correct semantic gesture.}
  \label{fig:qual}
\end{figure*}

The data contradiction results in two concrete failure modes in directly trained AT2M models:
\textbf{(i)~Condition signal overwhelmed by the prior.}
The strong unconditional motion prior
$p(x)$ dominates training and drowns the joint conditioning signal
$(a, c)$~\cite{ho2021cfg}: in the third row of Fig.~\ref{fig:qual}, when the input audio is enumerating \emph{``Thanksgiving, Christmas, the New Year''}, the corresponding semantic should be \emph{gesture for giving examples}. However, baseline methods only capture acoustic features in the audio and ignore the semantics.
\textbf{(ii)~Fixed condition strength at inference.}
The relative influence of $a$ and $c$ is baked into the trained weights; limited dataset scale makes models unable to  adjust to different control scale at inference time---e.g.\ to favour semantic fidelity over
beat alignment for a particular clip.

To tackle the challenges, our central observation is that strong A2M and T2M models already exist, each trained on its own large-scale A2M and T2M datasets.
The unimodal models already encode the full conditional structure needed for multimodal generation.  \textbf{AT2M capability can emerge through composition of unimodal models}.

\emph{Compositional diffusion}~\cite{liu2022compositional} offers a principled
escape from this data contradiction.
Rather than training a single model on scarce $(a, c, m)$ triples,
it trains each modality's expert on  large-scale unimodal corpus and
approximates $p(x \mid a, c)$ at inference time by composing
the experts' score predictions.

Previous works~\cite{du2023compositional, zhang2024energymogen} apply composition strategy by blending the experts'
score epsilon predictions linearly. The fusion works surprisingly well as a baseline, but harbours a structural flaw: it implicitly assumes that audio $a$ and text $c$ are conditionally independent given the noisy motion state, and thereby omits the \textbf{copula score}---the cross-modal dependence between $a$ and $c$ that neither marginal expert can represent.
We make this omission explicit in Section~\ref{sec:score} and derive the
exact correction term. This motivates \emph{CFGComposer}: a jointly-trained AT2M diffusion model
that delivers the interaction correction as an $\epsilon$-space residual, and applies compositional AT2M generation via copula score correction.

Our contributions can be summarized as follows:
\begin{enumerate}[leftmargin=*,itemsep=1pt,topsep=2pt]
  \item A rigorous derivation of the exact Product-of-Experts score
        decomposition for AT2M,
        establishing a principled probability-theoretic basis
        for compositional diffusion in the motion domain--going beyond prior
        EBM analogies by deriving the copula score $s_I$ exactly via the
        score identity without any conditional-independence approximation.
  \item \emph{CFGComposer}: a jointly-trained AT2M diffusion model that
        approximates the copula score via a CFG analogy and applies it
        as an $\epsilon$-space guidance correction, enabling
        flexible multimodal fusion via
        interaction scale $\lambda$ freely adjustable
        at inference.
  \item Empirical validation showing CFGComposer
        surpasses vanilla PoE fusion and joint-training baselines on both gesture semantic quality and audio beat alignment. Real-life humanoid deployment demonstrates high motion generation quality.
\end{enumerate}

\section{Related Work}
\label{sec:related}

\paragraph{Co-speech gesture generation}
Early systems map speech to gestures with hand-crafted
rules; later data-driven models learn the mapping
from text and audio~\cite{kucherenko2020gesticulator}, including end-to-end
training for a humanoid robot~\cite{yoon2019robots}.
Yoon~\emph{et al.}~\cite{yoon2020trimodal} are the first to condition on
the tri-modal context of text, audio, and speaker identity.
Diffusion models~\cite{tevet2022mdm,chen2024diffsheg,liu2024emage} trained
on BEAT~\cite{liu2022beat} and its SMPL-X release
BEATv2~\cite{liu2024emage} raise motion quality further, but leave gesture
semantics under-specified.
Recent AT2M methods add explicit semantic control through CLIP text
prompts~\cite{ao2023gesturediffuclip}, a separate semantic script
stage~\cite{zhi2023livelyspeaker}, a gated sparse semantic
layer~\cite{zhang2024semtalk}, speech-queried semantic
masks~\cite{zhang2025echomask}, LLM-based retrieval of canonical
clips~\cite{zhang2024sg}, or anti-inertia CFG masking for humanoid
robots~\cite{robogesture2026}.
All of them train one multimodal network on annotated tri-modal data or
rely on a curated motion library; CFGComposer aims to train only a lightweight residual estimator on a small paired subset to correct the composition of two independently trained unimodal experts.

\paragraph{Compositional diffusion and sampling correction}
Liu~\emph{et al.}~\cite{liu2022compositional} bring the product-of-experts
idea to diffusion models by summing the scores of independently trained
denoisers.
Their derivation assumes that the conditions are conditionally independent
given the sample.
At intermediate noise levels, however, the summed score is not the score of
the diffused product distribution.
Du~\emph{et al.}~\cite{du2023compositional} therefore correct the sampler
with annealed MCMC on the composed energy, and Feynman--Kac
correctors~\cite{skreta2025fkc} reweight a population of samples with
sequential Monte Carlo to target the true product.
In the motion domain, PriorMDM~\cite{shafir2024priormdm} composes a frozen
motion diffusion prior for long sequences, two-person interaction, and
body-part control.
EnergyMoGen~\cite{zhang2024energymogen} interprets the latent denoiser as
an implicit energy function equipped with conjunction and negation
operators.
These works either correct the mismatch through sampling or leave it
unmodelled.
CFGComposer instead learns the omitted copula score with a jointly
conditioned model, at the cost of one extra parallel forward pass per step.

\paragraph{Classifier-free guidance in multi-condition control}
Classifier-free guidance (CFG)~\cite{ho2021cfg} extrapolates from an unconditional
to a conditional prediction with a scalar weight.
In co-speech generation, FreeTalker~\cite{yang2024freetalker} trains one
diffusion model on heterogeneous speech-gesture and text-motion data and
composes the speech and text conditions with CFG at inference.
SynTalker~\cite{chen2024syntalker} exploits off-the-shelf T2M data through
multi-stage training and combines separately conditioned predictions at
inference in a separate-then-combine scheme.
Both works treat the guided difference as an empirical control process inside one
model.
CFGComposer uses the Product of Experts~(PoE) composition of frozen expert models as the baseline, then trains a residual model as an estimate of the copula
score, so $\lambda$ carries a probabilistic meaning rather than a tuning
heuristic.


\section{Problem Formulation}
\label{sec:formulation}

Let $x_0 \in \R^{T \times J}$ be a whole-body motion sequence of
$T = 300$ frames and $J = 60$ joints.
We are given:
\emph{(i)} an audio condition $a \in \R^{250 \times 512}$
(Whisper features at 50~fps for a 5-second clip);
\emph{(ii)} a text condition $c$ (gesture caption, e.g.\ ``thumbs up'');
\emph{(iii)} a gesture window $(g_s, g_e) \subset [0, T)$ specifying
where the semantic gesture should occur.
The goal is to sample from the joint conditional distribution
$p(x_0 \mid a,\,c)$ such that $x_0$ is rhythmically aligned with $a$
throughout and semantically consistent with $c$ within the gesture window.

\paragraph{Diffusion setup and pre-trained experts}
Currently, we already have well-trained unimodal experts, including audio-to-motion (A2M) and text-to-motion (T2M). Our unimodal experts are both in the DDPM~\cite{ho2020ddpm} framework
(cosine schedule, $x_0$-prediction, 1000 denoising steps) with
forward process $x_t = \at x_0 + \st\epsilon$, $\epsilon\sim\N(0,I)$,
and share the same~\cite{tevet2022mdm} Transformer backbone
with classifier-free guidance~\cite{ho2021cfg} ($p_{\mathrm{drop}}{=}0.1$).
$f_a$ is trained on $(x_0, a)$ pairs to model $p(x \mid a)$ (A2M);
$f_c$ is trained on $(x_0, c)$ pairs to model $p(x \mid c)$ (T2M).

Section~\ref{sec:theory} derives the exact score decomposition and a CFG-analogy estimate of the
copula score $s_I$ that PoE fusion omits.
Section~\ref{sec:methods} instantiates the PoE baseline
(Section~\ref{sec:vanilla}) and CFGComposer
(Section~\ref{sec:cfg-composer}), which adds the copula correction with a controllable weight $\lambda$.
Both experts are \emph{frozen} in all downstream use; Section~\ref{sec:pretrain} details their training procedures.

\section{Theoretical Analysis}
\label{sec:theory}

\subsection{Preliminary: Score Functions from Epsilon Networks}
\label{sec:prelim}

The DDPM training objective for a conditional model is:
\begin{equation}
  \mathcal{L}_\theta
  = \E_{x_0,\,t,\,\epsilon}\!\left[
      \bigl\|\epsilon_\theta(x_t,\,t,\,y) - \epsilon\bigr\|_2^2
    \right],
  \label{eq:ddpm-obj}
\end{equation}
where $x_t = \at x_0 + \st\epsilon$, $\epsilon \sim \N(0,I)$.
This objective is a form of \emph{denoising score matching}~\cite{vincent2011,song2019}:
at convergence, the learned predictor approximates the conditional score,
\begin{equation}
  \hat\epsilon_y(x_t, t)
  \;\approx\; -\st\,\nabla_{x_t}\!\log p_t(x_t \mid y),
  \label{eq:score-identity}
\end{equation}
where $p_t(x_t \mid y) = \int q(x_t \mid x_0)\,p(x_0 \mid y)\,dx_0$
is the noisy marginal under condition $y$~\cite{ho2020ddpm,song2021sde}.
We write $s_y(x_t, t) \!=\! \nabla_{x_t}\!\log p_t(x_t \mid y)$
for $y \in \{u,\,a,\,c,\,(a,c)\}$
($u$ denotes the unconditional model with condition dropped).

\subsection{The EBM Analogy for Compositional Inference}
\label{sec:ebm-prior}

A natural approach to inference-time composition interprets each
denoising network as an implicit energy-based model
(EBM)~\cite{lecun2006tutorial,du2019implicit}.
Following~\cite{liu2022compositional,zhang2024energymogen},
recall that an EBM defines a probability density via the
Boltzmann distribution:
\begin{equation}
  p_\theta(x) \;\propto\; \exp \bigl(-E_\theta(x)\bigr),
\end{equation}
where $E_\theta$ is a neural energy function.
Sampling proceeds by Langevin MCMC:
\begin{equation}
  x' = x - \eta\,\nabla_x E_\theta(x, y) + \mathcal{N}(0, \sigma^2 I).
  \label{eq:langevin}
\end{equation}
At each noise level $t$, the DDPM reverse step takes the same form,
with $\epsilon_\theta(x_t, t, y)$ in place of $\eta\,\nabla_x E_\theta$:
\begin{equation}
  x_{t-1} = x_t - \epsilon_\theta(x_t, t, y) + \mathcal{N}(0, \sigma_t^2 I).
  \label{eq:ddpm-step}
\end{equation}
Comparing at step size $\eta = 1$,
the denoising network approximates the energy gradient:
$\epsilon_\theta(x_t, t, y) \approx \nabla_x E_\theta(x_t, y)$.
This interprets each trained diffusion model as an implicit EBM
over the noisy state $x_t$.

\paragraph{Compositional inference via energy addition}
For a conjunction of conditions $a$ and $c$, the product of
Boltzmann distributions gives:
\begin{equation}
  p(x \mid a, c)
  \;\propto\;
  \frac{p(x \mid a)\,p(x \mid c)}{p(x)},
\end{equation}
i.e., energies add: $E(x; a, c) = E(x; a) + E(x; c) - E(x)$.
Substituting the EBM--diffusion correspondence and introducing
guidance weights $w_i \!\geq\! 0$ to temper each expert's contribution:
\begin{equation}
  \hat\epsilon_{\mathrm{PoE}}
  = \hat\epsilon_u + \sum_i w_i\bigl(\hat\epsilon_{c_i} - \hat\epsilon_u\bigr).
  \label{eq:ebm}
\end{equation}
Eq.~\eqref{eq:ebm} is empirically effective and widely
used~\cite{liu2022compositional,zhang2024energymogen}.
We call it the \emph{Product of Experts} (PoE) formula: the joint
distribution is the renormalized product of independent expert
distributions, $p(x|a,c) \propto p(x|a)\,p(x|c)/p(x)$.
We write $\hat\epsilon_{\mathrm{PoE}}$ for this prediction throughout.

\subsection{Exact Score Decomposition Reveals the Implicit Assumption}
\label{sec:score}

The EBM analogy provides an intuitive composition recipe. However, the sampling formula Eq.~\eqref{eq:ebm} is derived via an analogy from DDIM to EBM, rather than rigorous theory derivation. In particular, EBMs require that each $s_y = -\nabla_x E_y(x)$ be the gradient of some energy function (a conservative field), while the $\epsilon$ output of a neural network does not have this guarantee~\cite{salimans2021ebm}. Even if the sampling process is similar, the analogy is not precise.

We now ask what Eq.~\eqref{eq:ebm} implicitly assumes about the joint distribution $p(x \mid a, c)$, by deriving the composition formula using the score function.

At any noise level $t$, the joint log-density score can be decomposed:
\begin{align}
  \log p_t(x \!\mid\! a, c)
  &= \log p_t(x \!\mid\! a)
   + \log p_t(x \!\mid\! c)
   - \log p_t(x) \notag\\
  &\quad
  + \underbrace{
      \log \frac{p_t(x|a,c)\,p_t(x)}{p_t(x|a)\,p_t(x|c)}
    }_{I_t(x;\,a,c)}.
  \label{eq:logdensity}
\end{align}
Taking $\nabla_x$ of both sides:

\begin{proposition}[Exact Score Decomposition]
\label{prop:score}
For all\/ $x$, $t$, $a$, $c$, with no approximation:
\begin{equation}
  \boxed{
    s_{ac}
    = \underbrace{s_a + s_c - s_u}_{\textnormal{PoE baseline}}
    + \underbrace{s_I}_{\textnormal{copula score}},
  }
  \quad
  s_I = \nabla_x I_t(x;\,a,c).
  \label{eq:score-decomp}
\end{equation}
\end{proposition}

Here $I_t(x;a,c) = \log\frac{p_t(x|a,c)\,p_t(x)}{p_t(x|a)\,p_t(x|c)}$
is the \emph{log-copula density}: it captures the cross-modal dependence
between $a$ and $c$ that the product-of-marginals
$p_t(x|a)p_t(x|c)/p_t(x)$ cannot represent.
We call $s_I = \nabla_x I_t$ the \emph{copula score} for short.
$I_t$ vanishes---and $s_I = 0$---if and only if $a \perp c \mid x_t$.

\paragraph{EBM $=$ score decomposition under conditional independence}
Setting $s_I = 0$ in Proposition~\ref{prop:score}
(i.e., assuming $a \perp c \mid x_t$) and multiplying by $-\st$
via identity~\eqref{eq:score-identity} gives
$\hat\epsilon_a + \hat\epsilon_c - \hat\epsilon_u$,
which is exactly Eq.~\eqref{eq:ebm} at unit weights.
The EBM analogy therefore implicitly assumes
\emph{conditional independence at the noisy level},
and the copula score $s_I$ is precisely the correction
it omits: $\hat\epsilon_{\mathrm{PoE}} = \hat\epsilon_{ac} - \hat\epsilon_I$.

\paragraph{Why $s_I \neq 0$ at the noisy level}
Even if independence holds at the data level ($a \perp c \mid x_0$),
marginalising $x_0 \to x_t$ through the forward kernel $q(x_t|x_0)$
\emph{creates} dependence between $a$ and $c$ given $x_t$.
At high noise, $x_t$ is ambiguous and both conditions supply
complementary evidence about the denoising direction.

\subsection{Epsilon Parameterization and the Residual Target}
\label{sec:eps-param}

Using \eqref{eq:score-identity}, define optimal epsilon predictions
$\epsilon^*_y = -\st s_y$ for each $y$.
Applying this to Proposition~\ref{prop:score}:
\begin{equation}
  \epsilon^*_{ac}
  = \underbrace{\epsilon^*_a + \epsilon^*_c - \epsilon^*_u}_{\epsilon^*_{\mathrm{PoE}}}
    + \underbrace{\epsilon^*_I}_{\text{copula}},
  \quad \epsilon^*_I = -\st s_I.
  \label{eq:eps-decomp}
\end{equation}

\subsection{Tractable Approximation of $\hat{\epsilon}_I$}
\label{sec:approx}

The PoE baseline~\eqref{eq:ebm} sets $\epsilon^*_I = 0$
and absorbs the $\epsilon^*_u$ correction into per-model CFG scales.
From Eq.~\eqref{eq:eps-decomp}, this introduces a systematic bias
of exactly $\hat\epsilon_I$ relative to the true joint prediction
$\hat\epsilon_{ac}$: the PoE formula treats $a$ and $c$ as conditionally
independent given $x_t$, which is problematic in practice: audio and text are often not conditionally independent, 
the current semantic gesture is highly correlated with the spoken content.
This conditional-independence assumption prevents the model from
 achieving a balanced synchronization among multiple modalities. We therefore seek a tractable estimate of
 $\hat\epsilon_I$ to inject as a correction and recover 
 multimodal synchronization.

\paragraph{Estimating the guidance direction $\hat\epsilon_I$}
Proposition~\ref{prop:score} in $\epsilon$ space gives
$\hat\epsilon_I = \hat\epsilon_{ac} - \hat\epsilon_{\mathrm{PoE}}$,
and both terms on the right are estimated via the score
identity~\eqref{eq:score-identity}.
\emph{(i)} A jointly-trained AT2M model $f_\theta$, optimised on
paired $(x_0, a, c)$ data with CFG dropout, learns
$\hat\epsilon_{\mathrm{AT2M}}(x_t, t, a, c) \approx -\st s_{ac}$.
\emph{(ii)} The PoE baseline $\hat\epsilon_{\mathrm{PoE}}$ approximates
$-\st s_{\mathrm{PoE}}$, with the $-\hat\epsilon_u$ correction
absorbed into per-model CFG scales (Section~\ref{sec:eps-param}).
Subtracting (ii) from (i) under the score identity:
\begin{equation}
  \hat\epsilon_{\mathrm{AT2M}} - \hat\epsilon_{\mathrm{PoE}}
  \approx -\st(s_{ac} - s_{\mathrm{PoE}})
  = \hat\epsilon_I.
  \label{eq:cfg-analogy}
\end{equation}
This is structurally identical to standard CFG~\cite{ho2021cfg}:
$\hat\epsilon_{\mathrm{AT2M}}$ plays the role of the jointly-conditioned
prediction and $\hat\epsilon_{\mathrm{PoE}}$ the role of the baseline, with their difference estimating the guidance direction.

\paragraph{The CFGComposer formula}
Substituting~\eqref{eq:cfg-analogy} into the CFG template
gives the
complete composition formula:
\begin{equation}
  \hat\epsilon_{\mathrm{total}}
  = \hat\epsilon_{\mathrm{PoE}}
  + \lambda \cdot 
    \bigl(\hat\epsilon_{\mathrm{AT2M}} - \hat\epsilon_{\mathrm{PoE}}\bigr)
  \label{eq:cfg-composer-formula}
\end{equation}
where $\lambda \!\geq\! 0$ controls multimodal integration strength.
At $\lambda {=} 0$: PoE baseline (no cross-modal interaction);
at $\lambda {=} 1$: fully corrected toward the joint distribution
$p_t(x|a,c)$;
at $\lambda {>} 1$: the interaction is over-amplified, sharpening the
conditional response in the manner of strong CFG guidance.
Joint training of $f_\theta$ on paired $(x_0, a, c)$ data helps encode
the cross-modal interaction $s_I$ when $a$ and $c$ are
conditioned together, and independently-trained A2M or T2M models only need to 
recover the marginal conditionals $s_a$ and $s_c$.

\paragraph{Summary}
The PoE baseline~\eqref{eq:ebm} omits the copula score $s_I$ by
assuming $a \perp c \mid x_t$, which prevents it from capturing
cross-modal interaction between audio and text conditions.
The CFG-analogy estimate~\eqref{eq:cfg-analogy} recovers $\hat\epsilon_I$
from the score identity~\eqref{eq:score-identity}.
CFGComposer (Section~\ref{sec:cfg-composer}) operationalizes this
correction to achieve the multimodal synchronization between audio
and text that the PoE baseline cannot represent.

\section{Methods}

\label{sec:methods}
\begin{figure*}[t]
  \centering
  \includegraphics[width=0.96\textwidth]{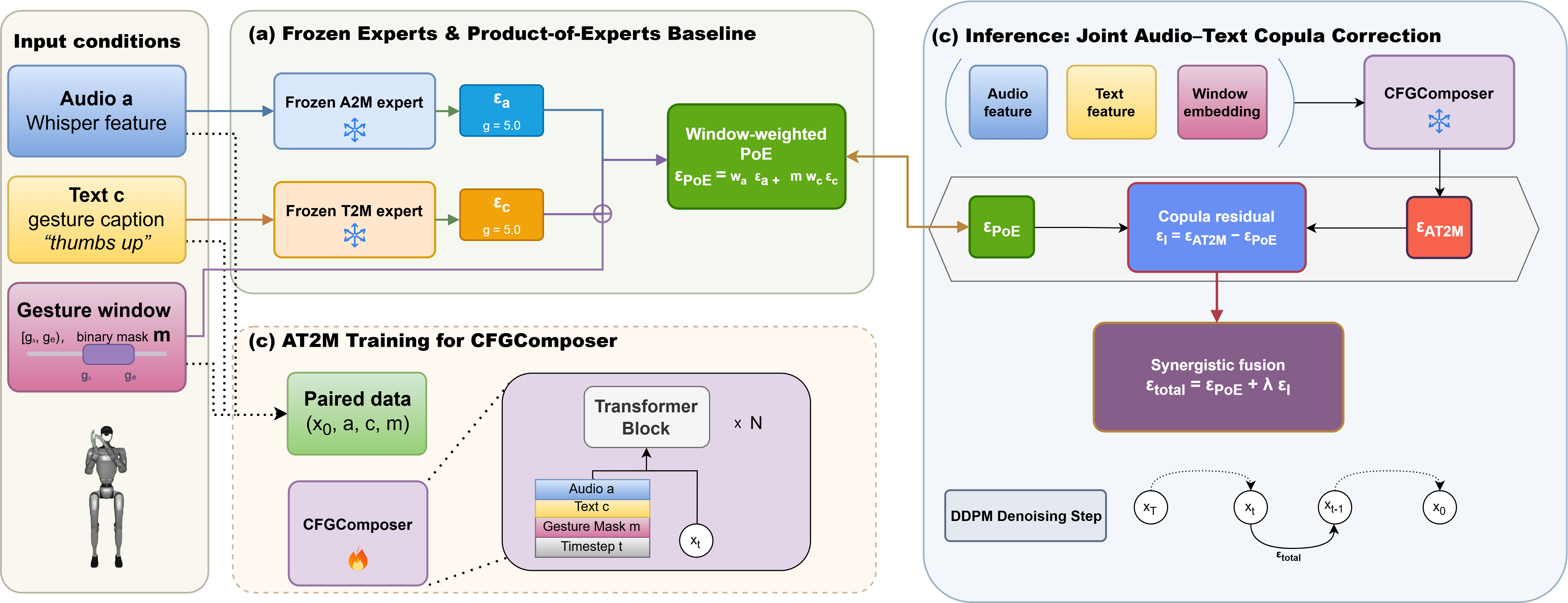}
  \caption{ \emph{CFGComposer} method overview.
    Pre-trained A2M and T2M experts compose at inference time via the
    exact PoE score decomposition (Eq.~\ref{eq:score-decomp}).
    CFGComposer trains a jointly-conditioned AT2M model and applies its
    CFG guidance direction as an $\epsilon$-space copula correction
    within the gesture window.}
  \label{fig:pipeline}
\end{figure*}

\subsection{Pre-training the A2M and T2M Experts}
\label{sec:pretrain}

Both expert models share the same~\cite{tevet2022mdm} Transformer backbone
($d{=}512$, 8 layers) and are trained separately
with the DDPM~\cite{ho2020ddpm} $x_0$-prediction objective
on BEAT~\cite{liu2022beat} with a cosine noise
schedule and CFG dropout
($p_{\mathrm{drop}}{=}0.1$):
\begin{equation}
  \mathcal{L}_{\mathrm{expert}}
  = \E_{x_0,\,t,\,y}\!\left[
      \bigl\|\hat{x}_0(x_t,\,t,\,y) - x_0\bigr\|_2^2
    \right].
  \label{eq:expert-loss}
\end{equation}

\textbf{A2M expert $f_a$}
Trained on $(x_0, a)$ pairs where $a\in\R^{250\times512}$ are
Whisper features (50\,fps).
A 2-layer Transformer audio encoder condenses the sequence to a single
conditioning token.

\textbf{T2M expert $f_c$}
Trained on $(x_0, c)$ pairs where $c$ is a gesture caption
(\emph{e.g.}, ``thumbs up'').
A frozen text encoder $g_c$ projects $c$ to a fixed-length conditioning
embedding.

\subsection{PoE Baseline}
\label{sec:vanilla}

The PoE baseline instantiates Eq.~\eqref{eq:ebm} with the two
pre-trained experts $f_a$ and $f_c$.
Each model is run with classifier-free guidance at scale $g {=} 5.0$:
\begin{equation}
  \hat\epsilon_y^g =  g\,\bigl(\hat\epsilon_y - \hat\epsilon_u\bigr),
  \quad y \in \{a,\,c\},
  \label{eq:cfg-expert}
\end{equation}
yielding CFG-scaled guidance $\hat\epsilon_a^g$ and $\hat\epsilon_c^g$.
The two experts are then combined per the PoE formula~\eqref{eq:ebm}
with frame-level weights: A2M contributes at $w_a {=} 1.0$ over all frames,
while T2M contributes at $w_c {=} 1.6$ only within the gesture window
$(g_s, g_e)$.
The combined prediction is normalised by the total weight:
\begin{equation}
  \hat\epsilon_{\mathrm{PoE}} = \begin{cases}
    \hat\epsilon_u + \dfrac{w_a \hat\epsilon_a^g + w_c \hat\epsilon_c^g}{w_a + w_c} & \text{if } i \in [g_s,\,g_e), \\[6pt]
    \hat\epsilon_u +\hat\epsilon_a^g & \text{otherwise,}
  \end{cases}
  \label{eq:vanilla-region}
\end{equation}
with a 15-frame smoothstep ramp at window edges to avoid discontinuities.

\subsection{CFGComposer}
\label{sec:cfg-composer}

\subsubsection{Training}
\label{sec:cfg-train}

CFGComposer trains a standard DDPM $x_0$-prediction model
on AT2M paired data $(x_0, a, c)$ with gesture-window mask $m$:
\begin{equation}
  \mathcal{L}_{\mathrm{CFG}}
  = \E\!\left[
      \bigl\|\hat{x}_0 - x_0\bigr\|_2^2
    \right],
  \label{eq:cfg-loss}
\end{equation}
where $\hat{x}_0 = f_\theta(x_t, t, a, c)$ is the model prediction.
Joint CFG dropout ($p_{\mathrm{drop}} {=} 0.1$) zeros both $a$ and $c$
simultaneously, enabling the unconditional path. 
Algorithm~\ref{alg:cfg-train} summarises training.

\begin{algorithm}[t]
\caption{CFGComposer Training}
\label{alg:cfg-train}
\begin{algorithmic}[1]
\Require Dataset $\mathcal{D}$,
         schedule $\{\bar\alpha_t\}$
\For{step $= 1, \ldots, 50\text{k}$}
  \State Sample $(x_0, a, c, m) \!\sim\! \mathcal{D}$;
         $t \!\sim\! \mathrm{Uniform}\{0,\!\ldots\!,\!999\}$
  \State $\epsilon \!\sim\! \N(0,I)$;
         $x_t \leftarrow \at x_0 + \st \epsilon$
  \State With prob.\ $0.1$: zero $a$ and $c$
         \Comment{joint CFG dropout}
  \State $\hat{x}_0 \leftarrow f_\theta(x_t,\,t,\,a,\,c,\,m)$
  \State $\mathcal{L} \leftarrow \|\hat{x}_0 - x_0\|_2^2$;
         update $\theta$
\EndFor
\end{algorithmic}
\end{algorithm}

\subsubsection{Architecture}
\label{sec:cfg-arch}

CFGComposer uses Transformer backbone
($d {=} 512$, 8 layers)
with dual conditioning of audio and text.
A 2-layer Transformer audio encoder maps raw Whisper features
$\in \R^{250 \times 512}$ to a single conditioning token;
text conditioning reuses the frozen T2M encoder's projected embedding to provide consistent text representations;
a per-frame window embedding
informs the model of the gesture-window layout.

\subsubsection{Inference}
\label{sec:cfg-infer}

At each denoising step, CFGComposer augments the PoE baseline prediction
with the copula correction $\hat\epsilon_I = \hat\epsilon_{\mathrm{AT2M}} - \hat\epsilon_{\mathrm{PoE}}$
(Eq.~\eqref{eq:cfg-composer-formula}).
The guidance scale $\lambda$ is the sole inference-time hyperparameter.
Algorithm~\ref{alg:cfg-infer} summarises inference.

\begin{algorithm}[t]
\caption{CFGComposer Inference}
\label{alg:cfg-infer}
\begin{algorithmic}[1]
\Require Pre-trained A2M $f_a$, T2M $f_c$; AT2M model $f_\theta$;
         schedule $\{\bar\alpha_t\}$; audio $a$, text $c$, gesture mask $m$;
         guidance scale $\lambda$
\State Sample $x_T \sim \N(0, I)$
\For{$t = T,\, T{-}1,\, \ldots,\, 1$}
  \State $\hat\epsilon_{\mathrm{PoE}} \leftarrow$
         Eq.~\eqref{eq:vanilla-region} with
         $f_a(x_t,t,a)$, $f_c(x_t,t,c)$
  \State $\hat{x}_0 \leftarrow f_\theta(x_t,\,t,\,a,\,c,\,m)$
         \Comment{AT2M forward pass}
  \State $\hat\epsilon_{\mathrm{AT2M}} \leftarrow (x_t - \at\hat{x}_0)/\st$
         \Comment{$x_0 \!\to\! \epsilon$ conversion}
  \State $\hat\epsilon_I \leftarrow \hat\epsilon_{\mathrm{AT2M}} - \hat\epsilon_{\mathrm{PoE}}$
         \Comment{copula correction}
  \State $\hat\epsilon_{\mathrm{total}} \leftarrow
         \hat\epsilon_{\mathrm{PoE}} + \lambda \cdot \hat\epsilon_I$
  \State $x_{t-1} \leftarrow \mathrm{DDPMStep}(x_t,\,\hat\epsilon_{\mathrm{total}},\,t)$
\EndFor
\State \Return $x_0$
\end{algorithmic}
\end{algorithm}

\section{Experiments}
\label{sec:exp}

\subsection{Experimental Setup}
\label{sec:setup}

\paragraph{Dataset}
All experiments use BEAT~\cite{liu2022beat}, a large-scale multimodal
co-speech corpus that pairs whole-body motion capture with synchronized
speech audio and semantic annotations (per-gesture text labels, emotional
tags, and event timestamps).
The dataset is retargeted into Unitree G1 with Brainco hand to allow native deployment on humanoid robots.

\paragraph{AT2M dataset construction}
BEAT's semantic annotations each specify a gesture window $(g_s, g_e)$
and we augment text captions $c$ (e.g.\ ``thumbs up'') for each annotation.
To build $(x_0, a, c, m)$ training triples, we extract a 5-second
(300-frame) motion clip around each annotation
together with the aligned audio sequence.
The gesture-window mask $m$ marks which frames fall inside $(g_s, g_e)$.

\begin{table*}[t]
\centering
\small
\setlength{\tabcolsep}{5pt}
\renewcommand{\arraystretch}{1.05}
\begin{tabular}{l rrrrrr rrrr}
\toprule
\multirow{2}{*}{Method}
  & \multicolumn{6}{c}{\textit{Automatic}}
  & \multicolumn{4}{c}{\textit{Human Evaluation}} \\
\cmidrule(lr){2-7}\cmidrule(lr){8-11}
& G-MSE$\downarrow$ & F-MSE$\downarrow$ & BA & SRGR$\uparrow$ & FID$\downarrow$ & DIV
& HN$\uparrow$ & SA$\uparrow$ & BM$\uparrow$ & HD$\uparrow$ \\
\midrule
Ground Truth
  & -- & -- & 0.1743 & 1.0000 & -- & 17.07
  & -- & -- & -- & -- \\
\midrule
DiffSHEG~\cite{chen2024diffsheg}
  & 0.0178 & 0.0163 & 0.2228 & 0.7749 & \underline{92.21}  &  8.82
  & $-0.089$ & \underline{$-0.132$} & $-0.125$ & $-0.147$ \\
SemTalk~\cite{zhang2024semtalk}
  & 0.0402 & 0.0411 & \textbf{0.1808} & 0.7354 & 177.91 & 11.28
  & $-0.269$ & $-0.137$ & $-0.257$ & $-0.197$ \\
Semantic Gesticulator~\cite{zhang2024sg}
  & 0.0254 & 0.0245 & 0.2127 & 0.7216 & 104.56 & \textbf{16.31}
  & $-0.196$ & $-0.146$ & $-0.183$ & $-0.171$ \\
RoboGesture~\cite{robogesture2026}
  & \underline{0.0159} & \underline{0.0138} & 0.1903 & \underline{0.7851} & 489.16 & 24.51
  & \underline{$-0.085$} & $-0.148$ & \underline{$-0.056$} & \underline{$-0.141$} \\
\midrule
\textbf{CFGComposer (Ours)}
  & \textbf{0.0079} & \textbf{0.0116} & \underline{0.1826} & \textbf{0.9916} & \textbf{27.18} & \underline{14.44}
  & $\mathbf{+0.648}$ & $\mathbf{+0.570}$ & $\mathbf{+0.630}$ & $\mathbf{+0.664}$ \\
\bottomrule
\end{tabular}
\caption{Comparison with AT2M methods on BEAT.
  Automatic metrics (left): $\downarrow$ lower is better, $\uparrow$ higher is better;
  GT row is the ground-truth reference.
  Human evaluation (right): average net preference score $(W{-}L)/N \in [-1,+1]$ from blind
  pairwise comparison, higher is better.
  \textbf{Bold}: best non-GT result per column. \underline{Underline}: second-best result.}
\label{tab:sota-auto}
\end{table*}

To compensate for the limited number of annotated gestures,
we augment each annotation with \emph{temporal-position variants}:
the same gesture window is embedded at 5 different frame offsets
within the 5-second clip, providing diverse audio-context for the same
semantic event.
The resulting AT2M dataset contains
775 training clips and 195 validation clips. We train our CFGComposer with 50k steps.

\paragraph{Pre-trained models}
A2M: 300k steps on the full BEAT audio-motion corpus.
T2M: 200k steps co-trained on gesture captions as text conditions and our collected motion-caption dataset.
Both use cosine DDPM with 1000 steps and $x_0$-prediction.

\paragraph{Automatic metrics}
\begin{itemize}[leftmargin=*,itemsep=1pt,topsep=2pt]
  \item \textbf{G-MSE}: Gesture-window MSE (de-normalized \texttt{qpos}).
        Primary metric for semantic fidelity.
  \item \textbf{F-MSE}: Full-clip MSE.
        Measures global motion quality.
  \item \textbf{BA}: Beat alignment (Bailando BC score)~\cite{siyao2022bailando}--measure alignment between audio onset and motion beats. 
        Closer to groundtruth is better.
  \item \textbf{SRGR}: Semantic-Relevant Gesture Recall~\cite{liu2022beat}---
        semantics-weighted PCK (Percentage of Correct Keypoints) between generated and ground-truth joint positions.
        We use gesture window as binary semantic relevance score $\lambda_s$ and $\delta = 0.1$ is the PCK threshold.
        Higher is better.
  \item \textbf{FID}: Fréchet Inception Distance from a pre-trained
        motion encoder. Lower is better.
  \item \textbf{DIV}: Motion diversity.
        Closer to groundtruth is better.
\end{itemize}

\paragraph{Human evaluation}
We conduct a blind pairwise comparison study with 17 evaluators on 50 test clips,
producing 500 comparison pairs (pairwise comparison  across 5 methods, 3{,}287 valid
judgements in total).
Evaluators watch two rendered avatar clips side-by-side and independently judge
four criteria: \textbf{Human-likeness \& Naturalness (HN)}, \textbf{Semantic Accuracy (SA)},
\textbf{Beat Matching (BM)}, and \textbf{Hand Detail (HD)}.
Method identities are hidden throughout. 
We report the \emph{net preference score} $(W{-}L)/N \in [-1,+1]$ per method,
where $W$ and $L$ are win and loss counts across all pairings and $N$ is total
comparisons; ties contribute 0 to both $W$ and $L$.

\subsection{Comparison with AT2M Methods}
\label{sec:comparison-sota}

We compare CFGComposer ($\lambda = 0.7$, see Section~\ref{sec:abl})
against four representative AT2M methods that also target the joint
audio-text-to-motion problem.

\paragraph{Baselines}
\textbf{DiffSHEG}~\cite{chen2024diffsheg}
generates holistic co-speech gestures via a diffusion model that conditions
jointly on speech audio and partial semantic cues.
\textbf{SemTalk}~\cite{zhang2024semtalk} decomposes motion generation into a
continuously-generated rhythmic base and a sparsely-activated semantic layer
fused via a learned semantic gate, directly targeting the beat-vs-semantic
trade-off.
\textbf{Semantic Gesticulator}~\cite{zhang2024sg} retrieves semantically
matching canonical gesture clips via an LLM-based generative retrieval model.
\textbf{RoboGesture}~\cite{robogesture2026} targets humanoid robot gesture
synthesis and introduces anti-inertia CFG masking to prevent modality eclipse.

\paragraph{Discussion}
Table~\ref{tab:sota-auto} reports automatic and human-evaluation metrics jointly;
the GT row serves as the reference baseline.
CFGComposer achieves the best G-MSE (0.0079), F-MSE (0.0116), SRGR (0.9916),
and FID (27.18) among all methods, which
indicates that composing
pre-trained experts preserves the natural gesture distribution and authentic to semantic labels.
DiffSHEG achieves the highest BA score (0.2228) by conditioning strongly on
audio, but its SRGR (0.7749) and FID (92.21) are substantially worse,
showing that heavy audio bias degrades semantic fidelity and realism.
RoboGesture generates high-diversity motions (DIV=24.51) but collapses
distributionally (FID=489.16), consistent with its data-augmentation which uses low-quality synthetic data as training data. 
SemTalk and Semantic Gesticulator both underperform on G-MSE and SRGR,
confirming that joint training on the scarce AT2M corpus is insufficient
for reliable semantic gesture generation.

\paragraph{Human evaluation}
CFGComposer is the only method with positive net preference scores across all
four dimensions (HN $+0.648$, SA $+0.570$, BM $+0.630$, HD $+0.664$),
winning $76$--$83\%$ of pairwise comparisons against every baseline.

\paragraph{Fine-tuning baselines on AT2M data}
A natural objection is that the baselines in Table~\ref{tab:sota-auto}
never saw our AT2M training set.
Table~\ref{tab:finetune} therefore fine-tunes the pretrained DiffSHEG and
RoboGesture on the same AT2M clips used by CFGComposer, either directly
or mixed with the BEAT training data.
Fine-tuning helps DiffSHEG modestly (G-MSE 0.0178 to 0.0139, SRGR 0.7749
to 0.7985) but leaves it far from CFGComposer (0.0079 and 0.9916).
RoboGesture improves in FID (489.16 to 179.52), as expected once the
target distribution is seen, and in SRGR (0.8018), yet G-MSE worsens and
DIV collapses to 0.01: the network memorises the few annotated gestures.
Mixing in BEAT data does not change this picture.
More AT2M supervision on a single joint network therefore does not close
the gap.

\begin{table}[t]
\centering
\footnotesize
\setlength{\tabcolsep}{3pt}
\renewcommand{\arraystretch}{1.02}
\resizebox{\columnwidth}{!}{%
\begin{tabular}{l rrrrrr}
\toprule
Variant & G-MSE$\downarrow$ & F-MSE$\downarrow$ & BA & SRGR$\uparrow$ & FID$\downarrow$ & DIV \\
\midrule
Ground Truth & -- & -- & 0.1743 & 1.0000 & -- & 17.07 \\
\midrule
DiffSHEG (pretrained)      & 0.0178 & 0.0163 & 0.2228 & 0.7749 & 92.21 & 8.82 \\
\quad AT2M fine-tune      & 0.0139 & 0.0118 & 0.1846 & 0.7985 & 68.45 & 11.18 \\
\quad BEAT+AT2M mixed     & 0.0154 & 0.0140 & 0.1838 & 0.7793 & 62.31 & 13.39 \\
\midrule
RoboGesture (pretrained)   & 0.0159 & 0.0138 & 0.1903 & 0.7851 & 489.16 & 24.51 \\
\quad AT2M fine-tune      & 0.0183 & 0.0179 & 0.2085 & 0.8018 & 179.52 & 0.01 \\
\quad BEAT+AT2M mixed     & 0.0177 & 0.0181 & 0.2530 & 0.6958 & 82.26 & 4.65 \\
\midrule
\textbf{CFGComposer (Ours)} & \textbf{0.0079} & \textbf{0.0116} & \textbf{0.1826} & \textbf{0.9916} & \textbf{27.18} & \textbf{14.44} \\
\bottomrule
\end{tabular}}
\caption{Fine-tuning baselines on our AT2M data.
  Pretrained rows repeat Table~\ref{tab:sota-auto}; metrics as there.
  \textbf{Bold}: best non-GT result per column.}
\label{tab:finetune}
\end{table}

\subsection{Ablation Studies on Guidance scale $\lambda$}
\label{sec:abl}

\begin{figure}[h]
  \centering
  \includegraphics[width=\columnwidth]{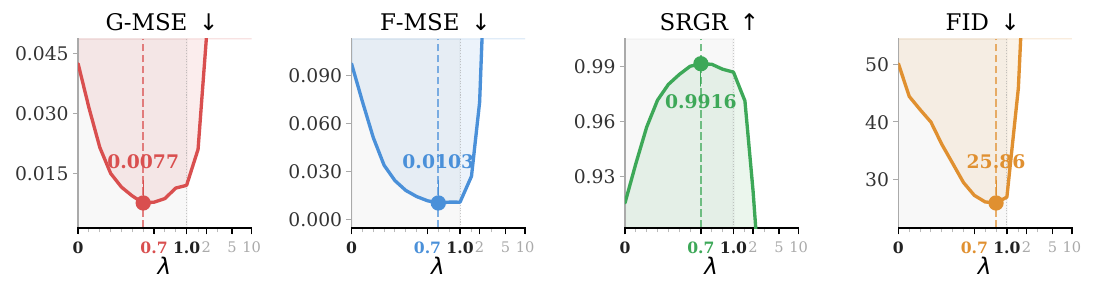}
  \caption{Four automatic metrics vs.\ guidance scale $\lambda$.
    X-axis: linear for $\lambda\!\in\![0,1]$ (shaded), log for $\lambda\!>\!1$.
    Annotated value: per-metric optimum; dashed line: chosen $\lambda{=}0.7$.}
  \label{fig:abl-lambda}
  \vspace{-3mm}
\end{figure}

Fig.~\ref{fig:abl-lambda} plots four automatic metrics over $\lambda \in [0, 3]$
(values beyond 3 degrade monotonically). We use the same model, only changing the guidance scale at inference.

The intermediate region $[0.6, 0.9]$ achieves the best balance:
composition with calibrated guidance allows the semantic signal to shape the
gesture window while the prior preserves motion fluency and beat alignment.
Each metric reaches its individual optimum in this region:
G-MSE at $\lambda{=}0.6$ (0.0077),
SRGR at $\lambda{=}0.7$ (0.9916),
F-MSE at $\lambda{=}0.8$ (0.0103),
and FID at $\lambda{=}0.9$ (25.86).
We select $\lambda{=}0.7$ as the operating point: it achieves the best SRGR
and is close to the per-metric optimum for G-MSE, F-MSE, and FID.
Beyond $\lambda{=}1$, over-amplification drives FID and G-MSE sharply upward
as the denoising trajectory leaves the natural motion manifold.

Two special cases deserve attention.
At $\lambda{=}0$ the correction vanishes and the method reduces to the
PoE baseline, which ignores the copula score and fails to coordinate audio
and text.
At $\lambda{=}1$ the method is functionally equivalent to deploying the
jointly-trained AT2M model alone; it underperforms $\lambda{=}0.7$ because
the semantic signal is overwhelmed by the strong motion prior, the
\textbf{data contradiction} of Section~\ref{sec:intro}.
A calibrated $\lambda$ therefore yields motions that are both rhythmically
synchronized and semantically aligned.

\subsection{Real-Robot Deployment}
\label{sec:robot}

To validate that CFGComposer transfers from simulation to a physical platform,
we deploy the full inference pipeline on a Unitree~G1 humanoid equipped
with BrainCo dexterous hands with 60 dof (7 root + 29 G1 + 24 Brainco Hands) in total.
Given a live speech utterance and a user-specified gesture caption,
CFGComposer generates each 5-second (300-frame) \texttt{qpos} clip in a
single reverse diffusion pass on a workstation (NVIDIA RTX~4090).
Each denoising step costs ${\approx}30$\,ms; with 50 DDIM steps the
first-clip latency is ${\approx}1.5$\,s, enabling real-time control.
The sequence is native to the G1's joint space, filtered for basic
self-collision, and streamed at 60\,Hz to a PD controller for the upper
body, while the Unitree~G1 locomotion policy handles lower-body balance.
BrainCo finger angles are recovered from the predicted finger joints and
sent to the hand controller over CAN bus.

\begin{figure}[t]
  \centering
  \includegraphics[width=0.48\textwidth]{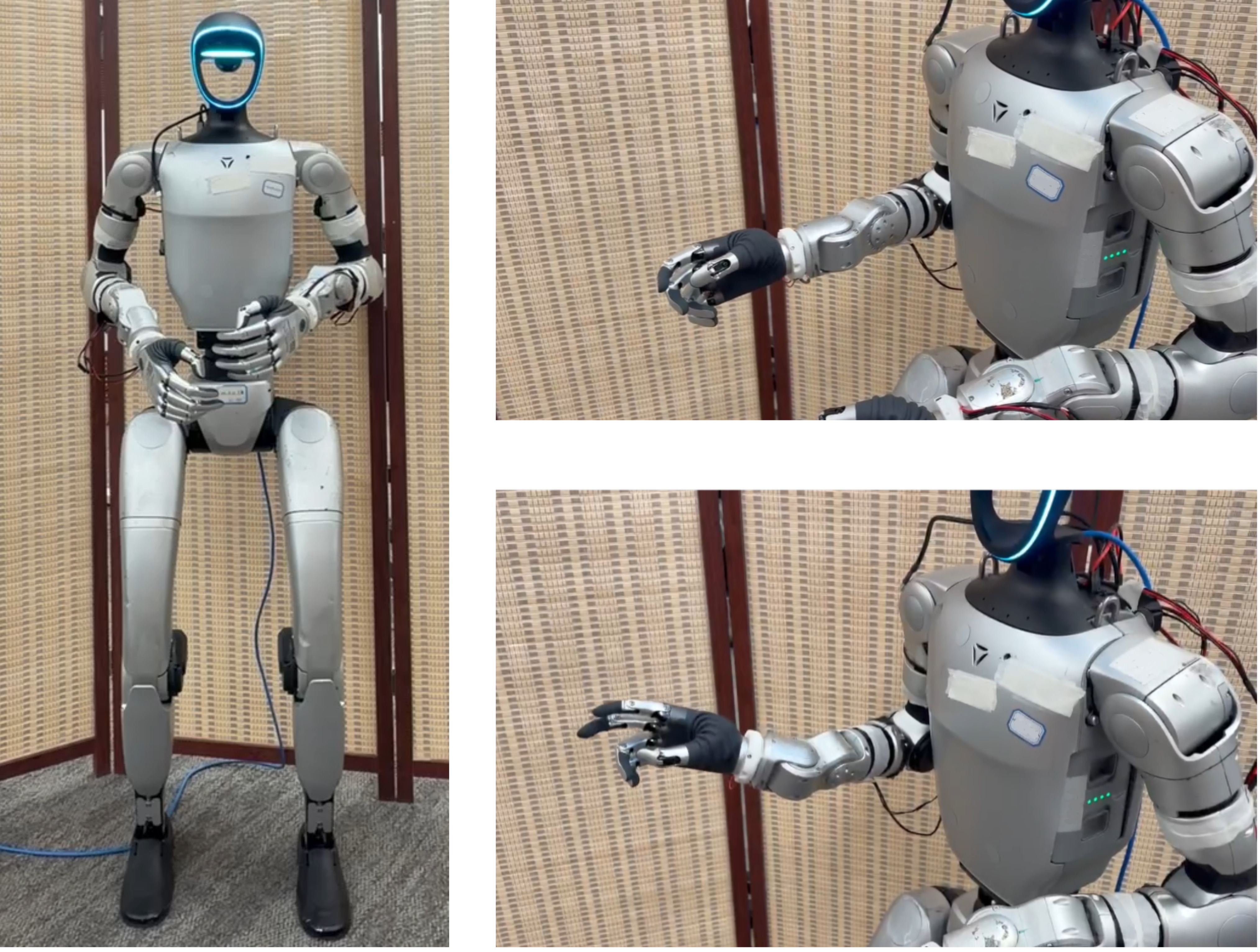}
  \caption{\textbf{Physical deployment on Unitree~G1 + BrainCo hand.}
    CFGComposer generates semantically accurate gestures synchronized
    to live speech. Generated results show agile finger movements.}
  \label{fig:robot}
   \vspace{-3mm}
\end{figure}

\paragraph{Qualitative observations}
Fig.~\ref{fig:robot} shows representative frames showing iconic gestures (counting two) with agile finger movements. CFGComposer reliably produces the target iconic shape within the
annotated gesture window while maintaining smooth, natural-looking
co-speech gestures.

\section{Conclusion}
\label{sec:conclusion}

We presented a theoretical and empirical study of compositional
audio-text-to-motion generation.
Our central finding is that the interaction term $s_I$ omitted
by vanilla Product-of-Experts fusion can be tractably estimated
via a CFG analogy: a jointly-trained AT2M model captures $s_{ac}$
while vanilla fusion captures $s_{\mathrm{PoE}}$, so their difference
in $\epsilon$ space approximates $\hat\epsilon_I$.
CFGComposer operationalises this insight, applying the correction
$\hat\epsilon_{\mathrm{total}} = \hat\epsilon_{\mathrm{PoE}}
  + \lambda \cdot (\hat\epsilon_{\mathrm{AT2M}} - \hat\epsilon_{\mathrm{PoE}})$
.
Physical deployment on a Unitree~G1 humanoid with BrainCo dexterous hands
confirms sim-to-real transfer without robot-specific fine-tuning.

\paragraph{Limitations and future directions}
The current system uses a global gesture caption for whole-body control;
body-part-level spatial conditioning would enable finer gestural control.
Extending the PoE decomposition to three or more conditions
(audio, text, speaker identity, emotion, etc.) via hierarchical 
inclusion-exclusion is a natural next step. For real-world deployment, it is possible to 
 combine motion generation with RL-based tracker to achieve more stable motion control, and optimize the diffusion sampling process to reduce human-robot interaction latency.

\end{document}